\documentclass[12pt]{article}

\usepackage{a4}
\usepackage{amsmath}
\usepackage{amssymb}
\usepackage{longtable}
\usepackage{psfig}
\usepackage{mathptm}

\begin{document}


\renewcommand{\baselinestretch}{1.08} \large\normalsize 
\renewcommand{\theequation}{\arabic{section}.\arabic{equation}}
\parskip2mm

{\hfill 
   \begin{tabular}{l}
     \today
   \end{tabular}
}

\vspace*{1.5cm}

\begin{center}

{\Large
 On the use of analytic expressions for the voltage distribution \\
 to analyze intracellular recordings. \\ \ \\
 {\it \large Note on ``{\it Characterization of subthreshold voltage fluctuations \\
 in neuronal membranes}''} \\
\ } \\

\vspace*{0.5cm}

{\bf M. Rudolph and A. Destexhe} \\

\vspace*{0.5cm}

{Unit{\'e} de Neuroscience Int{\'e}gratives et Computationnelles, CNRS, \\
91198 Gif-sur-Yvette, France.} \\

\vspace*{1cm}

\end{center}

\vspace*{3.5cm}
Corresponding author: \\
\\
\hspace*{2cm} Dr. A.\ Destexhe \\
\hspace*{2cm} Unit{\'e} de Neuroscience Int{\'e}gratives et
              Computationnelles (UNIC), CNRS \\
\hspace*{2cm} Bat.\ 32-33, 1 Avenue de la Terrasse \\
\hspace*{2cm} 91198 Gif-sur-Yvette \\
\hspace*{2cm} France \\
\\
\hspace*{2cm} Tel: 33-1-69-82-33-35 \\
\hspace*{2cm} Fax: 33-1-69-82-34-27 \\
\hspace*{2cm} e-mail: Destexhe@iaf.cnrs-gif.fr


\newpage

\begin{abstract}
\normalsize

Different analytic expressions for the membrane potential distribution of
membranes subject to synaptic noise have been proposed, and can be very helpful
to analyze experimental data.  However, all of these expressions are either
approximations or limit cases, and it is not clear how they compare, and which
expression should be used in a given situation.  In this note, we provide a
comparison of the different approximations available, with an aim to delineate
which expression is most suitable for analyzing experimental data.

\end{abstract}


\vspace{4mm}

\noindent Synaptic noise can be modeled by fluctuating conductances described by
Ornstein-Uhlenbeck stochastic processes (Destexhe, Rudolph, Fellous, \&
Sejnowski, 2001).  This system was investigated by using stochastic calculus to
obtain analytic expressions for the steady-state membrane potential (V$_m$)
distribution (Rudolph \& Destexhe, 2003; 2005).  Analytic expressions can also be
obtained for the moments of the underlying three-dimensional Fokker-Planck
equation (FPE) (Richardson, 2004), or by considering this equation under
different limit cases (Lindner \& Longtin, 2006).  One of the greatest promises
of such analytic expressions is that they can be used to deduce the
characteristics of conductance fluctuations from intracellular recordings {\it in
vivo} (Rudolph et al., 2004; 2005).  

A recent article (Lindner \& Longtin, 2006) provided an in-depth analysis
of some of these expressions, as well as different analytically-exact limit
cases.  One of the conclusions of this analysis was that the original expression
provided by Rudolph \& Destexhe (2003) was derived using steps that were
incorrect for colored noise, and that the expression obtained matches numerical
simulations only for restricted ranges of parameters.  The latter conclusion was
in agreement with the analysis provided in Rudolph \& Destexhe (2005).  Another
conclusion was that the ``extended expression'' proposed by Rudolph \& Destexhe
(2005), although providing an excellent fit to V$_m$ distributions in general,
does not match for some parameter values and in particular, it does not agree
with the analytically-exact static-noise limit.  This extended expression is
therefore not an exact solution of the system either.  Since several analytic
expressions were provided for the steady-state V$_m$ distribution (Rudolph \&
Destexhe, 2003; Richardson, 2004; Rudolph \& Destexhe, 2005; Lindner \& Longtin,
2006), and since all of these expressions are either approximations or limit
cases, it is not clear how they compare and which expression should be used in a
given situation.  In particular, it is unclear which expression should be used to
analyze experimental recordings.  In the present note, we attempt to answer these
questions by clarifying a number of points about some of the previous
expressions, and by providing a detailed comparison of the different expressions
available in the literature.

First, we would like to clarify a number of misleading statements we made
in the original article (Rudolph \& Destexhe, 2003), and which may lead to
confusion.  The goal of this paper was to obtain an analytic expression for the
steady-state V$_m$ distribution of membranes subject to conductance-based colored
noise sources.  To obtain this, we considered the full system under a $t \to
\infty$ limit.  In this limit, we noted that the noise time constants become
infinitesimally small compared to the time over which the system is considered,
and this property allowed us to treat the system as for white noise.  Our main
assumption was that this procedure would allow us to obtain the correct
steady-state properties like the V$_m$ distribution.  Our approach was to obtain
a simplified FPE which gives the same steady-state solutions as the FPE
describing the full system.  These assumptions were stated in the Results of
Rudolph \& Destexhe (2003), but were not clearly stated in the Abstract and
Discussion, and it could be understood that we claimed to provide an FPE valid
for the full system.  We clarify here that the treatment followed in that paper
did not intend to describe the full system, but was only restricted to
steady-state solutions.

Unlike the original expression (Rudolph \& Destexhe, 2003), which matches
only for a restricted range of parameters, the extended expression (Rudolph \&
Destexhe, 2005) matches for several orders of magnitude of the parameters (see
also supplementary information of Rudolph \& Destexhe, 2005).  Why the extended
expression matches so well, although it is not an exact solution of the system
(Lindner \& Longtin, 2006), is presently unknown.  It is not due to the presence
of boundary conditions, which could compensate for mismatches ``by chance''. 
Simulations with and without boundary conditions gave equally good fits for the
parameters considered here (see NEURON code in supplementary information).  Our
interpretation (Rudolph \& Destexhe, 2005) is that the $t \to \infty$ limit
altered the spectral structure of the stochastic process (filtering), and one can
recover a better spectral structure by following the same approximation for a
system that is solvable (e.g., that of Richardson, 2004) and correct it
accordingly.  Thus, as also found by Lindner \& Longtin (2006), the extended
expression is a very good approximation of the steady-state V$_m$ distribution. 
Other expressions have been proposed under different approximations (Richardson,
2004) or limit cases (Lindner \& Longtin, 2006) and also match well the
simulations for the applicable range of parameters.

Since different expressions were proposed corresponding to different
approximations (Rudolph \& Destexhe, 2003, 2005; Richardson, 2004; Lindner \&
Longtin, 2006), we investigated which expression must be used in practical
situations.  We have considered an extended range of parameters and tested all
expressions by running the model for 10,000 randomly-selected values within
this parameter space.  The results of this procedure are shown in Fig.~1A-D. The
smallest error between analytic expressions and numerical simulations was found
for the extended expression of Rudolph \& Destexhe (2005), followed by Gaussian
approximations of the same authors and that of Richardson (2004).  The fourth best
approximation was the static-noise limit by Lindner \& Longtin (2006).  By
scanning only within physiologically-relevant values based on conductance
measurements in cats {\it in vivo} (Rudolph et al., 2005), the same ranking was
observed (Fig.~1E), with even more drastic differences (up to 95\%; see
supplementary information).  Manual examination of the different parameter sets
where the extended expression was not the best estimate revealed that this
happened when both time constants were slow (``slow synapses''; decay time
constants $>$50~ms).  Indeed, performing parameter scans restricted to this region
of parameters showed that the extended expression, while still providing good fits
to the simulations, ranked first for less than 30\% of the cases, while the
static-noise limit was the best estimate for almost 50\% of parameter sets
(Fig.~1F; see details in supplementary information).  Scanning parameters within a
wider range of values including fast/slow synapses and weak/strong conductances
showed that the extended expression was still the best estimate (about 47\%),
followed by the static-noise limit (37\%; see supplementary information).

In conclusion, we have clarified here two main points.  First, we clarified
the assumptions and approximations that were too ambiguously stated in Rudolph \&
Destexhe (2003).  Second, we provided a comparison of the different expressions
available so far in the literature.  This comparison showed that, for
physiologically-relevant parameter values, the extended expression of Rudolph \&
Destexhe (2005) is the most accurate for about 80-90\% of the cases.  Outside of
this range, however, the situation may be different.  In systems driven by slow
noisy synaptic activity, the static-noise limit performed better.  We therefore 
conclude that, for practical situations of realistic conductance values and 
synaptic time constants, the extended expression provides the most accurate
alternative available.  This is also supported by the fact that the extended
was successfully tested in real neurons (Rudolph et al., 2004), which is perhaps 
the strongest evidence that this approach provides a powerful tool to analyze 
intracellular recordings.


\section*{Acknowledgments}

Research supported by CNRS and the Human Frontier Science Program.
Supplementary information can be found at http://cns-iaf.cnrs-gif.fr


\section*{References}

\begin{enumerate}

\small

\item[] Destexhe, A., Rudolph, M., Fellous, J.-M., \& Sejnowski, T.J. (2001).
  Fluctuating synaptic conductances recreate in vivo-like activity in
  neocortical neurons.
  {\it Neuroscience, 107}, 13-24.
\item[] Hines, M.L., \& Carnevale, N.T. (1997).
  The NEURON simulation environment.
  {\it Neural Comput., 9}, 1179-1209.
\item[] Lindner, B., \& Longtin, A. (2006).
  Comment on ``Characterization of subthreshold voltage fluctuations
  in neuronal membranes''
  {\it Neural Comput.,} (in press).
\item[] Richardson, M (2004).
  The effects of synaptic conductances on the voltage distribution
  and firing rate of spiking neurons.
  {\it Physical Review E69}, 051918.
\item[] Rudolph, M., \& Destexhe, A. (2003).
  Characterization of subthreshold voltage fluctuations
  in neuronal membranes.
  {\it Neural Comput., 15}, 2577-2618. 
\item[] Rudolph, M., \& Destexhe, A. (2005).
  An extended analytic expression for the membrane potential distribution
  of conductance-based synaptic noise.  
  {\it Neural Comput., 17}, 2301-2315.
\item[] Rudolph, M., Piwkowska, Z., Badoual, M., Bal, T., \&
  Destexhe, A. (2004).
  A method to estimate synaptic conductances from membrane
  potential fluctuations.
  {\it J. Neurophysiol., 91}, 2884-2896.
\item[] Rudolph, M., Pelletier, J-G., Par\'e, D. and Destexhe, A. (2005).
  Characterization of synaptic conductances and integrative properties during
  electrically-induced EEG-activated states in neocortical neurons in vivo.
  {\it J. Neurophysiol., 94}, 2805-2821.

\end{enumerate}



\clearpage

\renewcommand{\baselinestretch}{1.1} \large\normalsize

\begin{figure}[ht]
\centerline{\psfig{figure=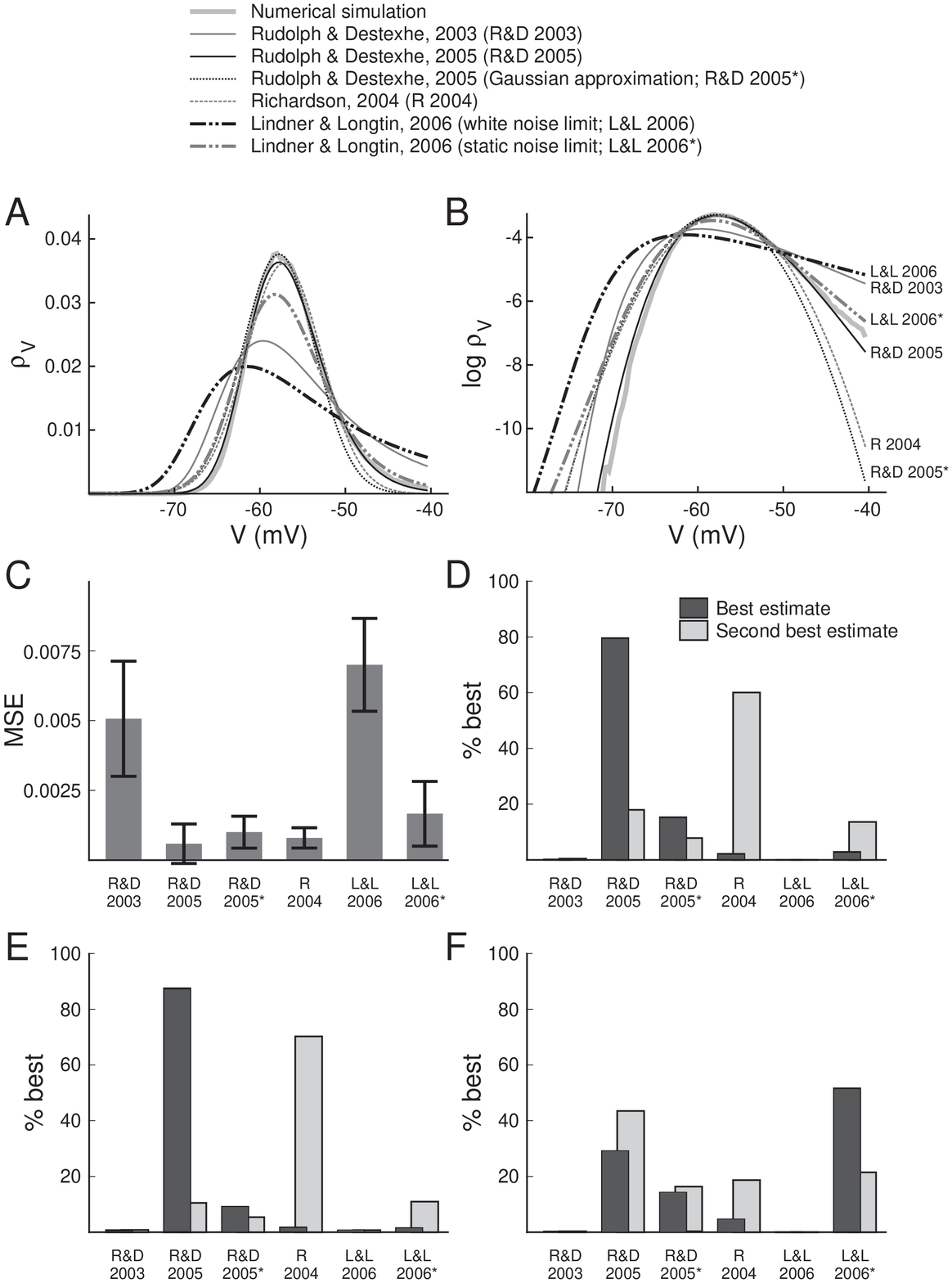,height=22cm}}
\end{figure}

\clearpage

{\bf Fig. 1}: \small Comparison of the accuracy of different analytic expressions
for the V$_m$ distributions of membranes subject to colored conductance noise. 
{\bf A.} Example of V$_m$ distribution calculated numerically (thick gray trace;
model from Destexhe et al., 2001, simulated during 100~s), compared to different
analytic expressions (see legend).  {\bf B.} Same as in A in log scale.  {\bf C.}
Mean square error obtained for each expression by scanning a plausible parameter
space spanned by 7 parameters.  10,000 runs similar to A were performed, using
randomly-chosen (uniformly distributed) parameter values.  For each run, the
mean-square error was computed between the numerical solution and each expression.
Parameters varied and range of values: membrane area $a$ =
5,000--50,000~$\mu$m$^2$, mean excitatory conductance $g_{e0}$ = 10--40~nS, mean
inhibitory conductance $g_{i0}$ = 10--100~nS, correlation times $\tau_e$ =
1--20~ms and $\tau_i$ = 1--50~ms.  The standard deviations ($\sigma_e$,
$\sigma_i$) were randomized between 20 and 33\% of the mean conductance values, to
limit the occurrence of negative conductances (in which case some analytic
expressions would not apply).  Fixed parameters: leak conductance density $g_L =
0.0452$~mS cm$^{-2}$ and reversal potential $E_L = -80$~mV, specific membrane
capacitance $C_m = 1~\mu$F cm$^{-2}$, and reversal potentials for excitation and
inhibition: $E_e = 0$~mV and $E_i = -75$~mV, respectively. {\bf D.} Histogram of
best estimates (black) and second best estimates (gray; both expressed in \% of
the 10,000 runs in B).  The extended expression ({\it R\&D 2005}) had the smallest
mean-square error for about 80\% of the cases.  The expression of Richardson
(2004) was the second best estimate for about 60\% of the cases.  {\bf E.}
Similar scan of parameters restricted to physiological values (taken from Rudolph
et al., 2005; $g_{e0}$ = 1--96~nS, $g_{i0}$ = 20--200~nS, $\tau_e$ = 1--5~ms and
$\tau_i$ = 5--20~ms).  In this case, {\it R\&D 2005} was the most performant for
about 86\% of the cases.  {\bf F.} Scan using strong conductances and slow time
constants ($g_{e0}$ = and $g_{i0}$ = 50--400~nS, $\tau_e$ and $\tau_i$ =
20--50~ms).  In this case, the static-noise limit {\it L\&L 2006*} was the most
performant for about 50\% of the cases.  All simulations were performed using the
NEURON simulation environment (Hines \& Carnevale, 1997) See supplementary
information for additional scans and the NEURON code of these simulations.


\clearpage


\section*{Appendix with Supplementary Information}

In this supplementary information, we provide more details about the
comparison between different analytic expressions for the
steady-state V$_m$ distribution of neurons subject to
conductance-based synaptic noise.  These different approximations are
respectively:

\begin{quote}
\begin{itemize}

\item[RD2003:] Original analytic expression of Rudolph \& Destexhe (2003);

\item[RD2005:] An ``extended'' analytic expression based on {\it RD2003},
where the time constants have been corrected to account for larger ranges of
parameters (Rudolph \& Destexhe, 2005);

\item[RD2005*:] A Gaussian approximation of the extended expression {\it RD2005}
(Rudolph \& Destexhe, 2005);

\item[R2004:] An effective time constant approximation (Richardson, 2004),
which is equivalent to a current-based approximation and is also Gaussian; 

\item[LL2006:] An analytically-exact white-noise approximation (limit of time
constants $\to$ 0; Lindner \& Longtin, 2006);

\item[LL2006*:] An analytically-exact static-noise approximation (limit of time
constants $\to \infty$; Lindner \& Longtin, 2006).

\end{itemize}
\end{quote}

Figure 1A-D of the paper shows a scan of 10,000 parameter values,
randomly chosen within reasonable bounds (larger than physiological
values).  For each parameter set, 100~sec of activity was simulated
and the V$_m$ distribution was computed numerically.  This numerical
estimate was then compared to each of the six expressions outlined
above.  In this scan, {\it RD2005} was the best estimate in about
80~\% of the cases, while the second-best estimate was {\it R2004} in
about 60~\% of the cases.

\subsection*{Additional analyses and scans of parameters}

In this supplementary information, we provide more examples of parameter scans
(using the same procedure as described in the paper), as well as illustrate some
typical situations.  As a first example, we scanned 10,000 parameter sets within
strictly ``physiological'' values.  Those values were obtained from a recent study
(Rudolph et al., 2005), in which the synaptic noise was analyzed from intracellular
recordings of neurons in cat parietal cortex {\it in vivo}.  This analysis used
both classic conductance analysis methods, the extended expression {\it RD2005}, as
well as direct matching of compartmental models to the recordings (see details in
Rudolph et al., 2005).  Both up/down states (Ketamine-Xylazine anesthesia) and
EEG-activated states were used for the analysis (n=12 cells).  The minimal and
maximal values for the conductances and variances obtained in those measurements
were used as bounds for choosing the 10,000 parameters.  The results of these
simulations are shown in Fig.~S-1A.  Similar to Fig.~1, {\it
RD2005} was the most accurate estimate for about 86~\% of the cases, followed by
the {\it R2004} approximation.  Because including two expressions biases the
analysis against {\it RD2005}, we also repeated the same analysis by removing the
Gaussian approximation {\it RD2005*}, as shown in Fig.~S-1B.  In this case, {\it
RD2005} was the best estimate for about 95\% of the parameter sets.

Manual examination of the cases for which {\it RD2005} was not the best estimate
revealed that this happened when both time constants were slow (``slow synapses'';
decay time constants $>$50~ms).  An example of such distribution is shown in
Fig.~S-2.   In this case, the static-noise limit {\it LL2006*} was the best
estimate, followed by {\it RD2005}.

To explore this region of parameters, we performed two additional runs of 5,000
randomly selected values of parameters, contrasting a region of parameter with fast
time constants with the same region with slow time constants.  When time constants
were fast, {\it RD2005}, {\it RD2005*} and {\it M2004} accounted for the best
performance (Fig.~S-3A), in agreement with above.  However, for slow time
constants, the most accurate estimate was obtained by using the static noise limit
(Fig.~S-3B; identical run as Fig.~1F of the paper).  The performance of static
noise limit is not surprising since this expression is specific for systems with
infinitely large noise time constants.

A last run was realized using a wider parameter range (Fig.~S-4), that included
physiological values, as well as slow synapses and strong conductances.  The
parameter space scanned included all regions of parameters scanned in all preceding
runs.  Based on a set of 10,000 parameter values randomly chosen within this
parameter space, the {\it RD2005} expression still provided the largest number of
best estimates (about 50\% of the cases), followed by the static-noise limit {\it
LL2006*} (37\%).  Similar values were obtained by removing {\it RD2005*} from the
analysis (Fig.~S-4B).

Based on these runs, we conclude that, for physiologically-relevant parameter
values, the extended expression {\it RD2005} is the most accurate for about 80-90\%
of the cases.  Outside of this range, however, the situation is different.  The
static noise limit can be a better approximation for systems with large noise time
constants (``slow synapses''), and should be used in such cases.

\subsection*{NEURON Code}

All simulations above and in the paper were done under the NEURON simulation
environment (Hines \& Carnevale, 1997).  The NEURON source code that was used for
the simulations shown here, as well as the code for data analysis and drawings,
can be found at the following location: \\
\underline{http://cns.iaf.cnrs-gif.fr/files/Note2006\_demo.zip} \\ This code
contains two parts.  First, a scanning program runs the numeric simulations for
the 10,000 parameters, and writes the results to a data file.  Second, an
analysis/drawing program reads this data file and creates the histograms shown in
Fig.~1.  The user can easily change the parameters and verify the simulations
shown here, or perform scans in unexplored parameter ranges, and thereby
contribute to a more rich analysis of how the different analytic expressions fit
numeric simulations.

Note that, contrary to the previous papers (Rudolph \& Destexhe, 2003, 2005), no
boundary conditions were used here, and the codes provided allow the conductance
to go negative.  Similar results were obtained when boundary conditions were used
(this is easy to modify in the code provided).

\subsection*{Experimental tests and analysis of experimental data}

Finally, another test of the analytic expressions is by comparing them directly
to experimental data.  The {\it RD2005} expression is the basis of a recently
proposed method to analyze intracellular recordings by fitting experimental
distributions, yielding estimates of parameters of the real synaptic noise, such
as the mean and variance of excitatory and inhibitory conductances (Rudolph et
al., 2004).  This method is presently used by several laboratories around the
world.  Related to the present paper, the {\it RD2005} expression was tested
against experimental data, in different ways.  First, the conductances obtained
by using the {\it RD2005}-based method were compared to other methods for
conductance analysis, as well as to the direct matching of computational models
to experimental data.   These different methods yielded consistent results for
activated states recorded intracellularly in cat parietal cortex {\it in vivo}
(see Rudolph et al., 2005), suggesting that {\it RD2005} is accurate for the
parameters corresponding to this type of synaptic noise in cortical neurons {\it
in vivo} (indeed those are the parameters shown in Fig.~S-1).

A second test, more severe, was realized using the dynamic-clamp technique.  The
synaptic noise produced spontaneously in ferret cortical slices (``up-states'')
was analyzed using {\it RD2005}, yielding estimates of the conductance
parameters.  An artificial synaptic noise was then generated using the estimated
parameters, and was re-injected in the {\it same neuron} during quiescent activity
using dynamic-clamp.  This yields a ``recreated'' state that can be compared to
the ``natural'' state.  This procedure was successful, as shown by the matching
of the natural and artificial V$_m$ distributions (see Fig.~7 and Fig.~8A in
Rudolph et al., 2004).  Another test, equally severe, was to first inject
synaptic noise with known parameters, and then compare the V$_m$ distribution
obtained in the real neuron with the analytic prediction of {\it RD2005}.  This
procedure also yielded consistent estimates (Fig.~8B in Rudolph et al., 2004).
 
These experiments and analyses show that the extended expression {\it RD2005} can
provide a very useful analysis tool for extracting conductances from experimental
data, and that the accuracy of this analysis is acceptable.  Other expressions
could possibly be used in similar paradigms, but this has not been done yet. 
Future experiments should be designed to address the respective accuracy of the
different expressions using similar procedures, which would constitute a further
test of their respective accuracy in physiological conditions.

\subsection*{Resources}

Electronic (PDF) copies of the paper and supplementary information are available
at: \\ \underline{http://cns.iaf.cnrs-gif.fr/files/Note2006.pdf} \\
\underline{http://cns.iaf.cnrs-gif.fr/files/Note2006\_suppl.pdf}

The NEURON code corresponding to the simulations is available at: \\
\underline{http://cns.iaf.cnrs-gif.fr/files/Note2006\_demo.zip}


\clearpage

\renewcommand{\baselinestretch}{1} \large\normalsize

\begin{figure}[ht]
\centerline{\psfig{figure=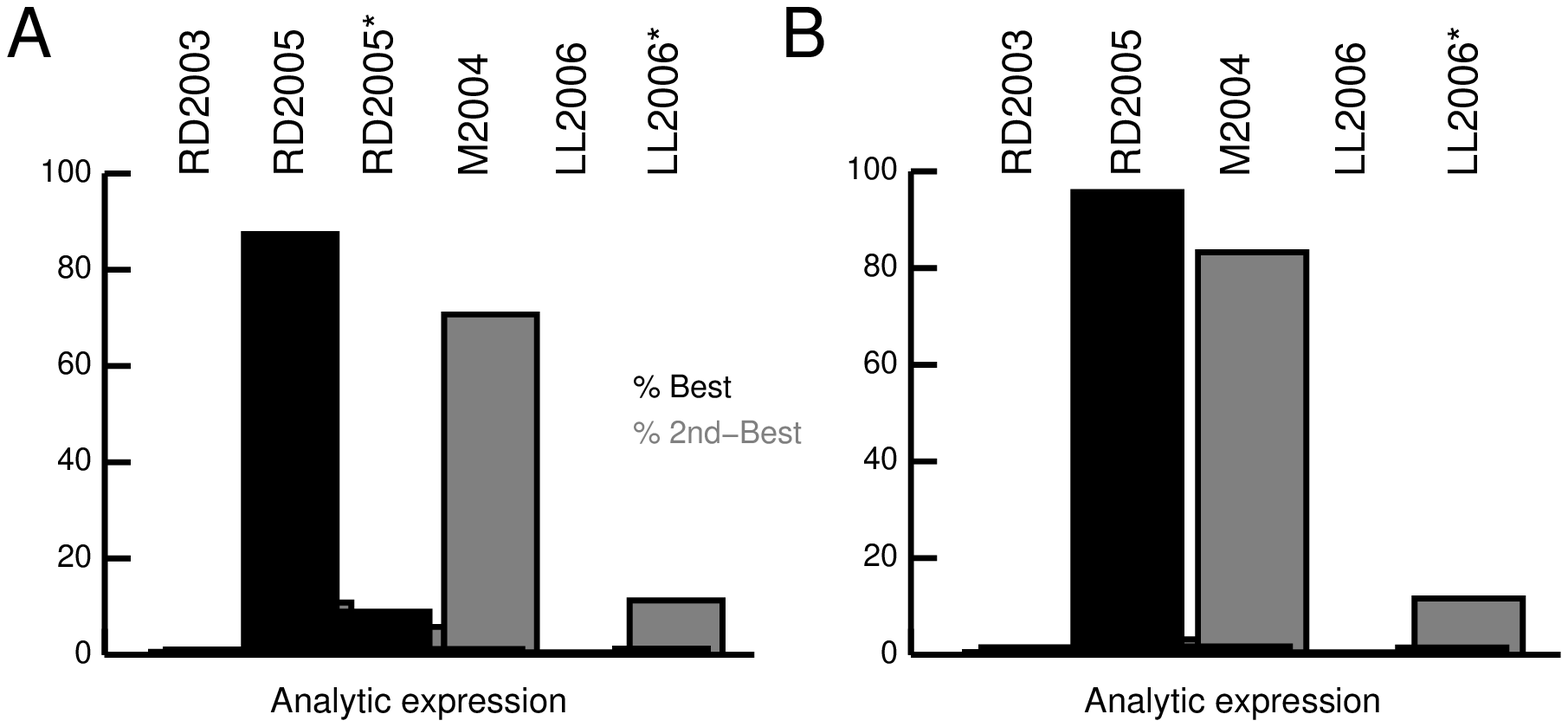,width=16cm}}
\end{figure}

{\bf Fig. S-1}: \small Histogram of best estimates for physiological values of
parameters. {\bf A.}  Additional scan of 10,000 runs of parameters using
randomly-chosen parameter (same procedure as in Fig.~1 of the accompanying article)
within the following range: membrane area $a$ = 5,000--50,000~$\mu$m$^2$, mean
excitatory conductance $g_{e0}$ = 1--96~nS, mean inhibitory conductance $g_{i0}$ =
20--200~nS, correlation times $\tau_e$ = 1--5~ms and $\tau_i$ = 5--20~ms.  The red
dashed histograms show the second best estimates.  The extended expression ({\it
RD2005}) had the smallest mean-square error for about 86\% of the cases.  {\bf B}. 
Same set of simulations, but the histogram was calculated by removing {\it
RD2005*}.  In this case, {\it RD2005} was the most accurate for about 95\% of the
cases.



\clearpage

\renewcommand{\baselinestretch}{1} \large\normalsize

\begin{figure}[ht]
\centerline{\psfig{figure=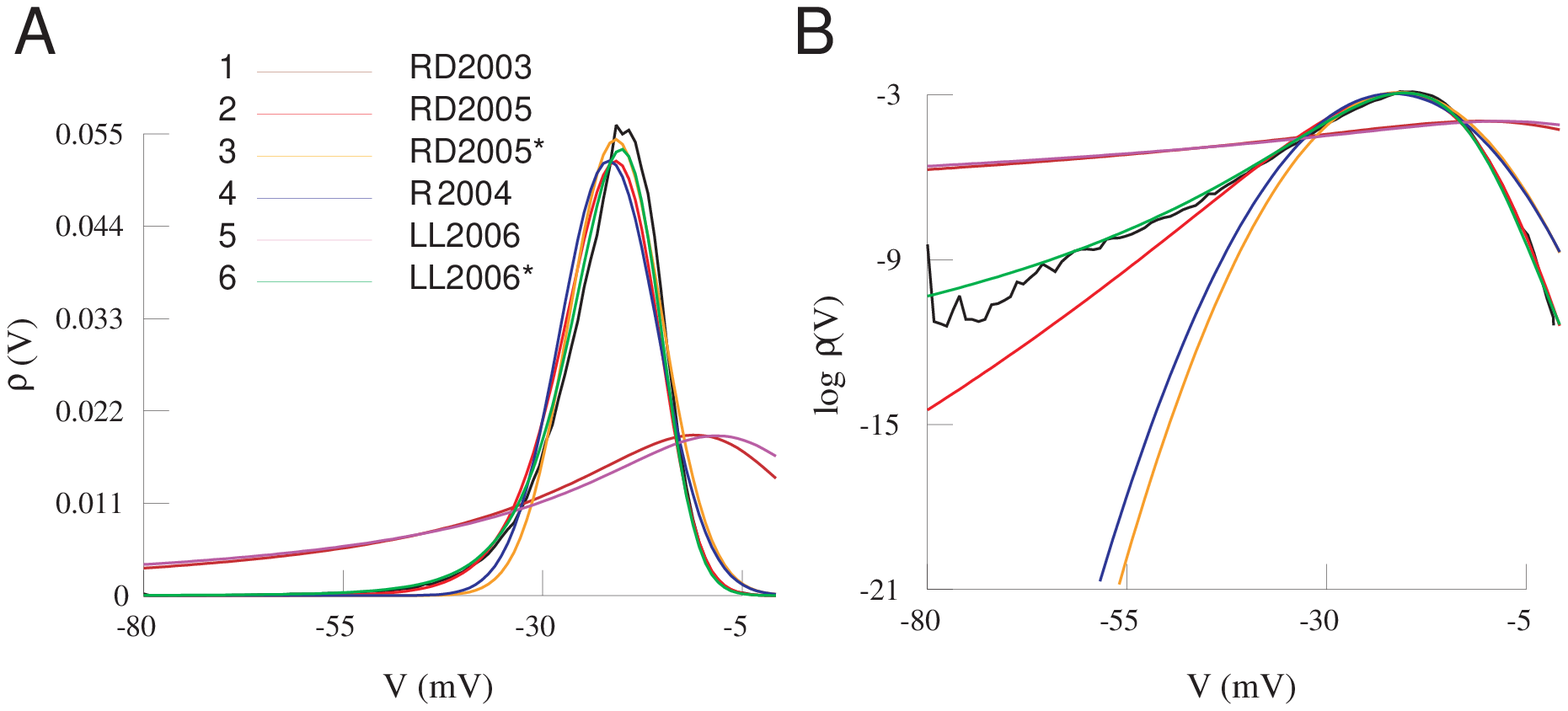,width=16cm}}
\end{figure}

{\bf Fig. S-2}: \small Example of V$_m$ distribution for parameters where the
static noise limit is the best approximation.  The V$_m$ distributions are shown
using a similar layout as Fig.~1A-B of the accompanying article (left: linear
scale, right: log-scale; color code in inset).  The best fit was in this case the
static noise limit ({\it LL2006*}, green), while {\it RD2005} was second best
(red).  Parameters: membrane area $a$ = 37286~$\mu$m$^2$, excitatory conductance
$g_{e0}$ = 400~nS, $\sigma_e$ = 130~nS, mean inhibitory conductance $g_{i0}$ =
141~nS, $\sigma_i$ = 39~nS, correlation times $\tau_e$ = 35.4~ms and $\tau_i$ =
20.8~ms.



\clearpage

\renewcommand{\baselinestretch}{1} \large\normalsize

\begin{figure}[ht]
\centerline{\psfig{figure=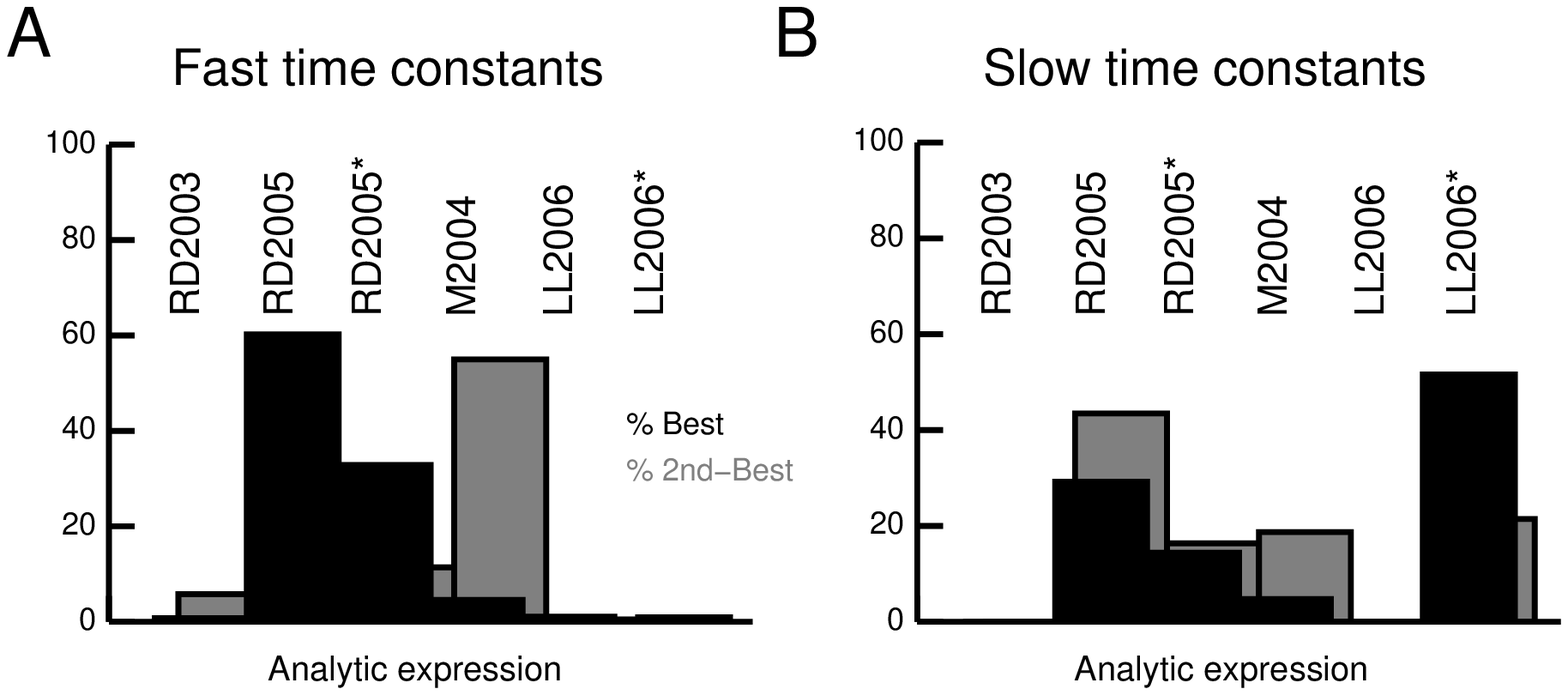,width=16cm}}
\end{figure}

{\bf Fig. S-3}: \small Histogram of best estimates for fast and slow time
constants.  Two additional scans of 5,000 parameters each are shown in {\bf A} and
{\bf B}, using the same procedure as in Fig.~1 of the accompanying article.  The
same parameters were used in both scans ($a$ = 5,000--50,000~$\mu$m$^2$; $g_{e0}$ =
1--50~nS, $g_{i0}$ = 1--50~nS), except for the time constants ($\tau_e$ = 1--5~ms
and $\tau_i$ = 5--20~ms in {\bf A}; $\tau_e$ and $\tau_i$ = 50--200~ms in {\bf B}).
The red dashed histograms show the second best estimates.  For fast time constants,
{\it RD2005} was the most accurate estimate for about 60\% of the cases, whereas
for slow time constants, {\it LL2006*} was more accurate for about 50\% of the runs.



\clearpage

\renewcommand{\baselinestretch}{1} \large\normalsize

\begin{figure}[ht]
\centerline{\psfig{figure=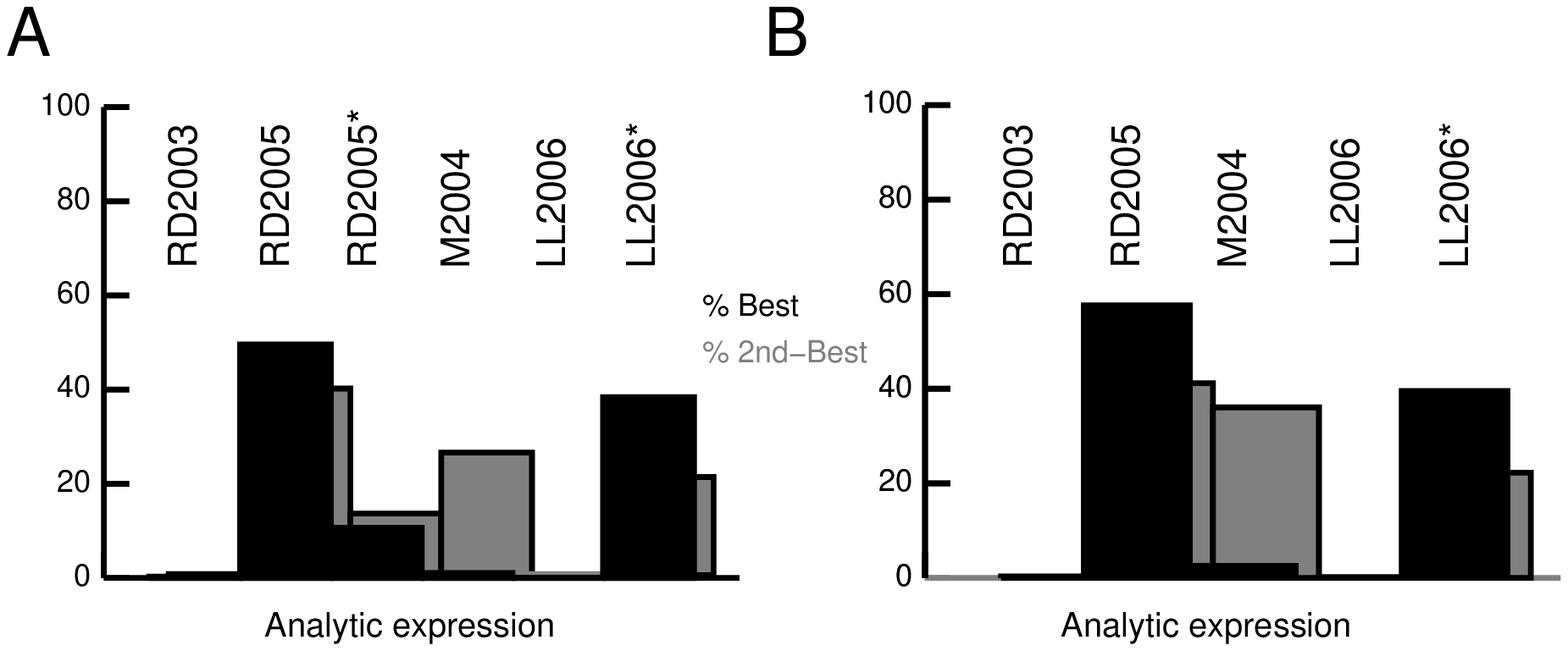,width=16cm}}
\end{figure}

{\bf Fig. S-4}: \small Histogram of best estimates for scans within a wide range of
parameter values. {\bf A.}  Additional scan of 10,000 runs of parameters using
randomly-chosen parameters (same procedure as in Fig.~1 of the accompanying
article) within the following range: $a$ = 1,000--100,000~$\mu$m$^2$, $g_{e0}$ =
1--300~nS, $g_{i0}$ = 1--300~nS, $\tau_e$ = 1--200~ms and $\tau_i$ = 1--200~ms. 
The red dashed histograms show the second best estimates.  The extended expression
({\it RD2005}) had smallest mean-square error for about 50\% of the cases.  {\bf
B}.  Same set of simulations, but the histograms were calculated by removing {\it
RD2005*}.  In this case, {\it RD2005} was the most performant for about 57\% of the
cases.


\end{document}